\documentclass[10pt,a4paper,oneside,onecolumn]{article}

\usepackage{ICROMA}
\setcitestyle{numbers,sort&compress}
\usepackage{enumerate}
\usepackage{amsmath,amssymb}
\usepackage{graphicx}
\usepackage{float}
\usepackage{caption}
\usepackage{subcaption} 
\usepackage{algorithm}
\usepackage{algpseudocode}
\usepackage{tikz} 
\usepackage{easyReview}
\usetikzlibrary{shapes.geometric, arrows}

\title{Track Component Failure Detection Using Data Analytics over existing STDS Track Circuit data}
\author{
        López Francisco\textsuperscript{*}, 
        Di Santi Eduardo\textsuperscript{*}, 
        Lefebvre Clément{1}, 
        Mijatovic Nenad,\\
        Pugnaloni Michele, 
        Martín Victor, 
        Saiah Kenza
}
\affiliation{
         Digital and Integrated Systems, Alstom\\
\noindent
E-mails: \href{mailto:francisco.lopez@alstomgroup.com}{francisco.lopez@alstomgroup.com}, 
\href{mailto:eduardo.di-santi@alstomgroup.com}{eduardo.di-santi@alstomgroup.com}, 
\href{mailto:clement.lefebvre-renard@alstomgroup.com}{clement.lefebvre-renard@alstomgroup.com}, 
\href{mailto:nenad.mijatovic@alstomgroup.com}{nenad.mijatovic@alstomgroup.com}, 
\href{mailto:michele.pugnaloni@alstomgroup.com}{michele.pugnaloni@alstomgroup.com}, 
\href{mailto:jonathan.brown@alstomgroup.com}{jonathan.brown@alstomgroup.com}, 
\href{mailto:victor-andres.martin@alstomgroup.com}{victor-andres.martin@alstomgroup.com}, 
\href{mailto:kenza.saiah@alstomgroup.com}{kenza.saiah@alstomgroup.com}
}

\begin{document}

\maketitle

\begin{abstract}
    Track Circuits (TC) are the main signalling devices used to detect the presence of a train on a rail track. It has been used since the 19th century and nowadays there are many types depending on the technology. As a general classification, Track Circuits can be divided into 2 main groups, DC (Direct Current) and AC (Alternating Current) circuits. This work is focused on a particular AC track circuit, called “Smart Train Detection System” (STDS), designed with both high and low-frequency bands.  This approach uses STDS current data applied to an SVM (support vector machine) classifier as a type of failure identifier. The main purpose of this work consists on determine automatically which is the component of the track that is failing to improve the maintenance action. Model was trained to classify 15 different failures that belong to 3 more general categories. The method was tested with field data from 10 different track circuits and validated by the STDS track circuit expert and maintainers. All use cases were correctly classified by the method.
\end{abstract}

\keywords{
    Track circuit, STDS, machine learning, SVM, maintenance
}

\section{Introduction}
\begin{figure}[H]
    \centering
    \includegraphics[width=0.75\linewidth]{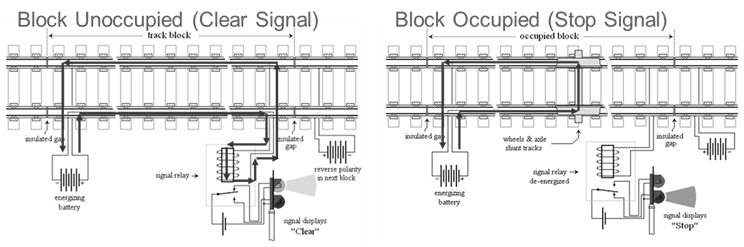}
    \caption{Track circuit behaviour schema}
    \label{fig:track_circuit_schema}
\end{figure}

A track circuit is an electrical system that detects the presence of a train on the tracks by passing a current through the rails, which acts as a conductor. In its initial form, track circuits consisted of a battery and a relay with adjustable resistors to set the transmitted signal gain and receiver operating point.  Sections of track are electrically isolated by insulated joints in each rail.  The transmitted signal travels through a single rail, through the relay at the opposite end, then returning to the transmitter through the other rail. Track circuits follow the closed loop principle, which means that any failure results in the safest state (track occupied) as shown in Figure~\ref{fig:track_circuit_schema}. Because of this, track circuits also provide detection of broken rails.  

Nowadays, there are many types of track circuits. The last state of the art ones provide enhanced performance, integrating sophisticated signalling systems to improve operation and safety. 

Track-circuit failures have an important impact as they imply a stop of operations and an economic impact for both the railway operator and its customers \cite{train_derailments}. In the past century, all data generated by track circuits was being discarded. When a failure occurred, it was necessary to stop the operations, send a maintenance team to the field, collect some data on-site to diagnose the problem and fix the failure if possible. Nowadays, thanks to the current technological advances, data can be stored by suppliers and be used for multiple purposes. One of them is the predictive maintenance of track circuits, which improves the overall reliability of railway networks and reduces the burden of curative actions in case of failures. 

The work presented in this publication focuses on Smart Train Detection System (STDS) track circuit. The main purpose is to use STDS current and voltage data to understand which component of the track circuit is failing, leading to a quicker maintenance action. The main advantage compared with other detection methods as \cite{network_rail_1}, \cite{network_rail_2}, \cite{acoustic}, \cite{image} is that there is no need to deploy any extra sensor or device to the field or doing a dedicated inspection since this approach is using data that is being already generated by the track circuit itself. Some studies such as \cite{nenad_1}, \cite{nenad_2}, already use data available on the field without extra hardware.
A list of different failures ordered by frequency of occurrence is provided. The objective is to detect those failures by inspecting Voltage and Current signal patterns and inform which component of the track is failing.

Machine learning and deep learning algorithms are typical solutions for those kind of use cases. Multiple studies as \cite{survey}, \cite{svm_survey}, \cite{svm_based_framework}, \cite{fault_detection_nn}, \cite{nn_in_railway}, \cite{severity_evaluation}  show promising results for those kind of analysis.\\
The following approach is made of an SVM classifier (Support Vector Machine) as a type of failure identifier.

\section{Methodology}
\subsection{ Data Preprocessing}
The data available is a field dataset which contains data of 10 different track circuits. Each file covers a period of 1 month of data sampled every 1 second and contains the received RMS voltage (root mean square voltage) which is equivalent to the direct current voltage of an alternating current source. It is a steady signal that has a fixed value over time (it can vary of  0.5v due to hardware design constraints) whose value depends on the configuration of the track circuit but stays within the range of [19v, 21v] when the track is free. If the track is occupied, the voltage value drops to 0. There is a fixed defined threshold of 17v that indicates if the track is occupied (below 17v) or free (above 17v). The track circuit works in a nominal mode if it is working under these conditions. 

\begin{figure}[H]
    \centering
    \includegraphics[width=0.75\linewidth]{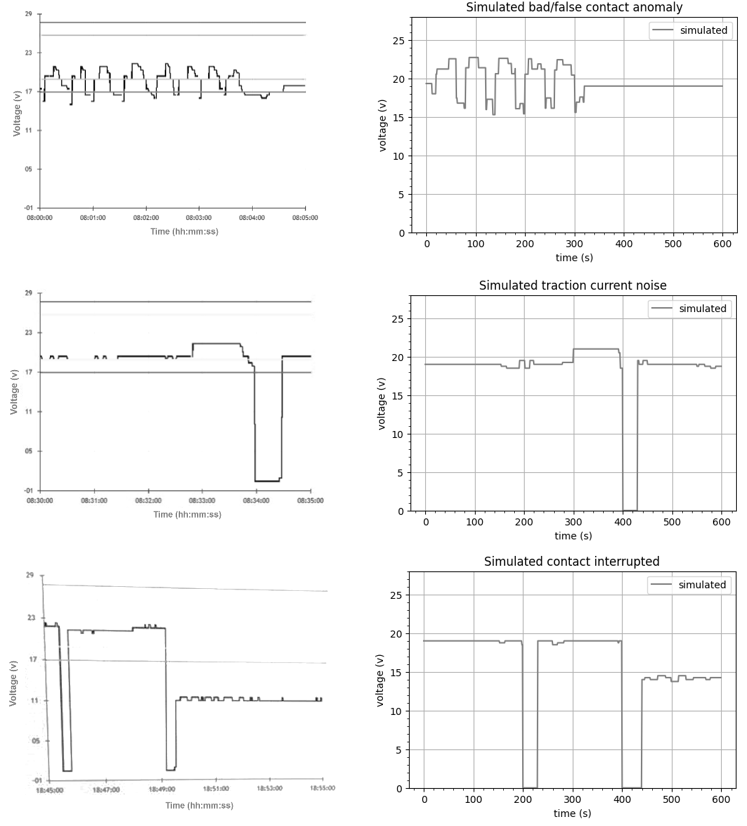}
    \caption{Anomalies. Data from the field (left),  Simulated samples (right). From top to bottom, the first graph corresponds to bad/false contact anomaly, the second to traction current noise anomaly and the third one to contact interrupted anomaly }
    \label{fig:anomalies_example}
\end{figure}

If any of these conditions is altered, the track circuit is considered to be working in an anomalous status. This work focuses on the 3 following anomalies, see Figure~\ref{fig:anomalies_example}: 
\begin{enumerate}
    \item \textit{Bad/false contacts:} The voltage value starts to oscillate up and down and can cross the occupation threshold, causing an intermittent false occupancy in the track. This behaviour can last for several minutes and can be due to a terminal box, transformer or resistor tightening loose, an inductive case tightening loose or a bad contact of the impedance bond.
    \item \textit{Traction current noise:}  The voltage increases before a train passes. It can be due to an unbalance of the track circuit or saturation of the inductive connections.
    \item \textit{Contact interrupted:} The voltage drops after a train passes through. It can drop below the occupancy threshold, causing a “permanent” false occupancy on the track. It usually happens when the train breaks a physical component of the track, causing for example a TC power cable interruption, detachment of the terminal box cable or a braid detachment of the transmission terminal box.
\end{enumerate}

The difference between the last two anomalies (Traction current noise and Contact interrupted)  lies on if the failure occurs before or after a train passes the track. The traction current noise anomaly is related to some noise that can be induced into the track due to the proximity of a train. The anomalous behaviour should disappear when the train is far away and the signal should come back to normal values. There can be a time window before this behaviour disappears depending on the length of the track and the distance of the train. The contact interrupted anomaly is a problem that appears after a train crosses the track, meaning that the problem has been produced by the train itself and the anomalous behaviour is going to be permanent on time.

\subsection{Failure Generator}
  It would be impossible to train a classifier without a labelled dataset, so an failure generator was developed following the indications of the expert to create as many samples as needed. Examples from field vs simulated anomalies are provided in Figure~\ref{fig:anomalies_example}:
\begin{enumerate}
    \item \textit{Bad/false contact anomaly:} Looking at Figure~\ref{fig:anomalies_example}, the signal can be decomposed into two simpler ones. One with a very large period, of 1 to 2 minutes (to simulate big changes), and then another one with a higher frequency and smaller amplitudes (to simulate small steps). The resulting signal is the sum of those two signals:
    \begin{enumerate}
        \item First, create a square signal with a frequency between 0.005 and 0.03 Hz. This means that the signals have a period between 30s and 200s and an amplitude between 0 or 4 volts.
        \item Second, add a "noise" signal with a variation in amplitude between -0.5 to 0.5 volts.
    \end{enumerate}	
    These parameters can be modified in order to make the anomaly more or less severe. 
    \item \textit{Traction current noise:} From a steady signal set to a value between 19v and 21v plus some noise, train passes can be simulated by setting the signal value to 0. To simulate a traction current noise, one can increase the voltage over the default value before the signal drops to 0.
    \item \textit{Contact interrupted:} As for a traction current noise, a normal signal is created with some noise and an occupancy. To simulate a contact interrupted, one can decrease the voltage below the normal value after the signal recovers from 0.
\end{enumerate}

\subsection{Modelling}
The following approach aims at correctly classify the anomalies described in the previous sections.

\subsubsection{SVM for classification}
\begin{figure}[H]
    \centering
    \includegraphics[width=1\linewidth]{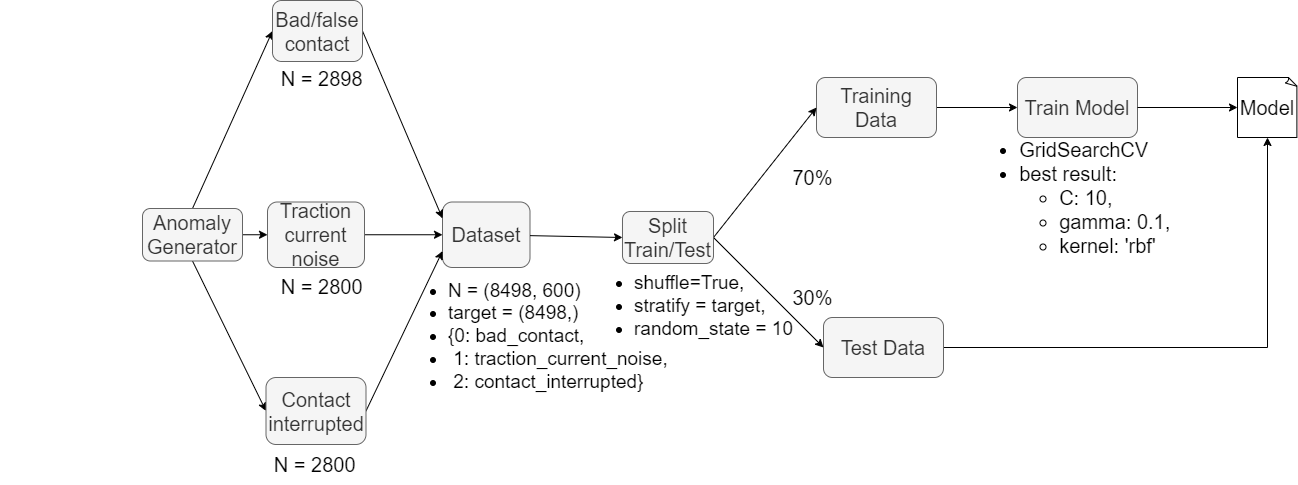}
    \caption{Classifier training steps}
    \label{fig:anomaly_classifier_training}
\end{figure}

An SVM (Support Vector Machine) is used as a failure classifier. The intuition behind the algorithm is the following. In 2 dimensions, one can separate 2 regions of a sheet just drawing a line. If this is extended to 3 dimensions, this line will convert into a plane that divides the space into 2 regions. Extending this idea to more dimensions, the "line" that divides the space is called an hyper-plane. The algorithm tries to maximize the distance between the boundary and the closest points of each region to this boundary (which are called support vectors). These boundaries can be more complex than linear, so the algorithm presents a kernel parameter. The C  parameter (common to all SVM kernels) trades off misclassification of training examples against simplicity of the decision surface and gamma parameter can be seen as the radius of influence of the support vectors selected \cite{statistical_learning}.
To train it, the failure generator is used to create the training set (See Figure~\ref{fig:anomaly_classifier_training}). Received voltage is selected as target feature and reshaped to have samples containing 600 points. Looking at Figure~\ref{fig:anomalies_example}, one can see that the anomalies have a magnitude order in time of 5-10 minutes. Hence, data is sampled every 10 minutes. Since the sample rate is one sample/s, the input size of the model is  600 points per sample. ~2800 samples of each anomaly category are created. Training and test datasets are created using a split of 70\% \ and 30\% respectively. The model is trained doing a gridSearch with the following values for C and gamma parameters: 
C:[0.1,1,10,100],
gamma:[0.0001,0.001,0.1,1],
kernel:['rbf','poly']. 
The best result obtained is using C: 10, gamma: 0.1, kernel: 'rbf'

\section{Results}
\begin{figure}[H]
    \centering
    \includegraphics[width=0.9\linewidth]{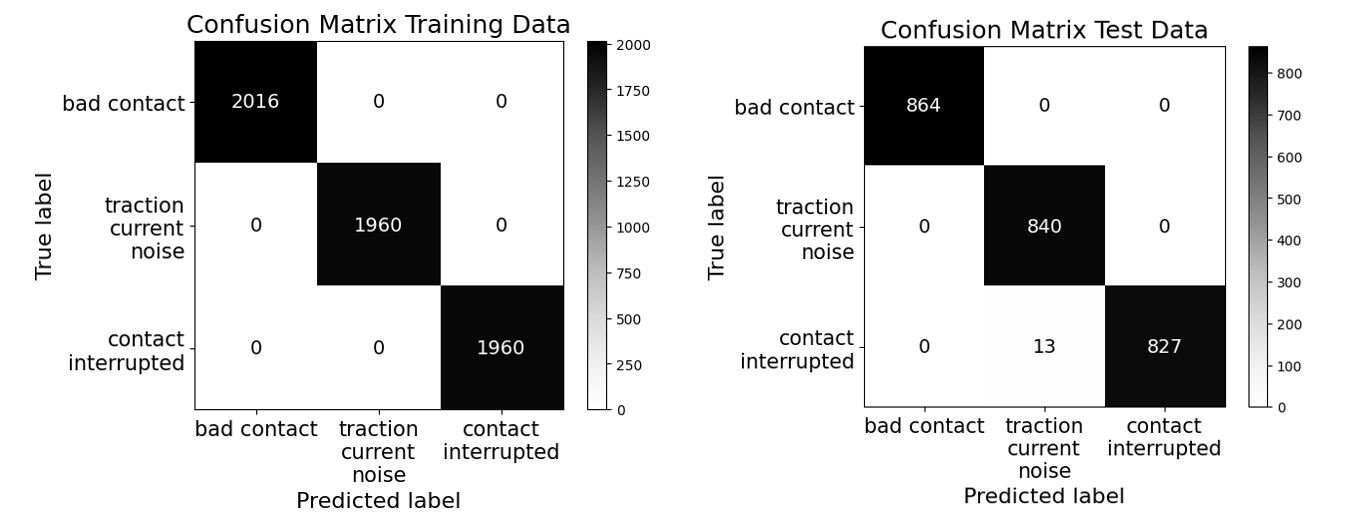}
    \caption{Confusion matrices. Training set (left), Test set (right)}
    \label{fig:confusion}
\end{figure}

Figure~\ref{fig:confusion} shows the confusion matrices of the classifier for the training and test phases. As one can see on the figure, the model has an average precision of 99.4\% on the test data. There are just some samples labelled as contact interrupted that were classified as a traction current noise. This means that only in the ~1\% of the cases, the maintainer will check a component of the track that is not failing. It is important to remark that these results are obtained using laboratory data. 
 
\section{Preliminary Estimation of Large-Scale Implementation Benefits}

The proposed failure detection system for track circuits has the potential to bring significant benefits when implemented at a large scale. This section provides a preliminary estimation of the expected advantages in terms of maintenance efficiency, cost savings, and overall railway operations. Given the variability in track circuit deployments and failure rates, these figures should be considered as a rough estimation based on available data.
\subsection{Reduction in Track Circuit Failures and Downtime}
Implementing this system across a railway network enables early detection of component failures, reducing unexpected track circuit failures. Failures in track circuits often lead to service disruptions, causing delays and economic losses.

According to available failure data, track circuit failures account for approximately \textbf{15--30\%} of all signalling-related failures in railway networks. In the UK, Network Rail reports around \textbf{6,000 track circuit failures per year}. If STDS track circuits represent \textbf{5--10\%} of total track circuits, a rough estimation suggests that STDS-related failures could account for \textbf{~600 failures per year in a large railway network} (e.g., UK, France, Italy). 

Extrapolating to a \textbf{European-wide scale} (\textit{10 major networks}), we estimate \textbf{~6,000 failures per year}, and over a \textbf{10-year period}, this amounts to \textbf{~60,000 failures}.

\textbf{Preliminary estimation:} If predictive maintenance reduces unscheduled failures by \textbf{30--50\%}, this could result in an estimated \textbf{15--25\% reduction in operational delays}, preventing \textbf{thousands of service disruptions annually}.

\subsection{Cost Savings in Maintenance}
Traditional maintenance strategies rely on reactive interventions, requiring emergency site visits and manual inspections. By enabling predictive failure detection, maintenance teams can proactively intervene, optimizing resources and minimizing unplanned interventions.
\textbf{Preliminary estimation:} A reduction of unnecessary site visits and emergency maintenance interventions by \textbf{20--40\%} could lead to an estimated \textbf{10--20\% reduction in overall maintenance costs} for railway operators.
\subsection{Increased Track Availability and Service Reliability}
Proactive failure detection enhances track availability and ensures more reliable railway operations. By preventing unnecessary disruptions, train schedules can be maintained more efficiently.
\textbf{Preliminary estimation:} Improving track availability by even \textbf{1--2\%} could result in substantial operational efficiency gains and increased revenue retention for large railway networks.
\subsection{Scalability and Ease of Implementation}
A key advantage of this approach is that it leverages existing track circuit data without requiring additional sensor installations or hardware modifications. Given that most modern railway systems already use track circuits, the deployment of this method would primarily require software integration.

\textbf{Preliminary estimation:} Since \textbf{80--90\% of existing railways use some form of track circuit technology}, large-scale adoption would be feasible with minimal infrastructure modifications, making this solution practical and cost-effective.

\section{Conclusion}
This approach demonstrated to work with consistent classification rates. It is extendable, new failures (confirmed by the expert) can be easily added to the classifier. The main advantage compared with other approaches is that it does not need to deploy any additional sensor or dedicated field inspection to collect the data.

As a future work, it would be needed to augment the labelled dataset with field data to improve the test phase of the classifier.

\section*{Author Contributions}
{Francisco López and Eduardo Di Santi contributed equally to this work as co-first authors.}


\end{document}